\documentclass[a4paper,12pt]{article} 

\usepackage{geometry,epsfig,subfigure}
\usepackage{amsmath}
\usepackage{amssymb}
\usepackage{graphics,epsfig,mydef}
\usepackage[square,sort,comma,numbers]{natbib}

\usepackage[usenames]{color}
\usepackage{fancyhdr,url}   
\usepackage{setspace}  
\usepackage{multirow}   
\usepackage{rotating}    
\usepackage{colortbl}     
\usepackage{relsize} 
\usepackage{booktabs}
\usepackage{cancel}
\usepackage{enumerate}
\numberwithin{equation}{section}
\usepackage{soul} 
\usepackage{xcolor} 
\usepackage{pdflscape}
\usepackage{hyperref}
\usepackage[capitalise]{cleveref}
\usepackage{lineno}
\usepackage{graphicx}
\usepackage{etoolbox}
\usepackage{epstopdf}

\definecolor{darkblue}{rgb}{0,0,0.8}
\definecolor{darkgreen}{rgb}{0,0.5,0}

\long\def\symbolfootnote[#1]#2{\begingroup \def\thefootnote{\fnsymbol{footnote}}\footnote[#1]{#2} \endgroup} 

\newcommand{\degree}{\ensuremath{ ^{\circ}  }}

\graphicspath{{figures/}}

\usepackage{amsthm}

\newcommand{\HRule}{\rule{0.9\linewidth}{0.2mm}}

\begin{document}
\renewcommand*{\thepage}{\arabic{page}}

\setstretch{1.3}

\begin{center}
\large
\textbf{The effect of disjoining pressure on the shape of condensing films in a fin-groove corner\\}

\normalsize
\vspace{0.2cm}
Osman Akdag$^{a}$, Yigit Akkus$^{a}$, Zafer Dursunkaya$^{b}\symbolfootnote[1]{e-mail: \texttt{refaz@metu.edu.tr}}\!$\\
\smaller
\vspace{0.2cm}
$^a$ASELSAN Inc., 06172 Yenimahalle, Ankara, Turkey\\
$^b$Department of Mechanical Engineering, Middle East Technical University, 06800 \c Cankaya, Ankara, Turkey\\
\vspace{0.2cm}
\end{center}

\begin{center} \noindent \HRule \\ \end{center}
\vspace{-0.6cm}
\begin{abstract}

\noindent 

Thin film condensation is commonly present in numerous natural and artificial processes. Phase-change driven passive heat spreaders such as heat pipes, which are widely used in electronics cooling, employ a continuous condensation process at the condenser region. When the wick structure of a heat pipe is composed of grooves, the top surfaces of the walls (fins) located between consecutive grooves function as the major source of condensation and the condensate flows along the fin top into the grooves. Modeling of this condensation problem is vital for the proper estimation of condensation heat transfer, which constitutes the basis for the overall performance of an heat pipe together with the evaporation process. In the current study, a solution methodology is developed to model the condensation and associated liquid flow in a fin-groove system. Conservation of mass and momentum equations, augmented Young-Laplace equation and Kucherov-Rikenglaz equation are solved simultaneously to calculate the film thickness profile. The model proposed enables the investigation of the effect of disjoining pressure on the film profile by keeping the fin-groove corner, where the film becomes thinnest, inside the solution domain.  The results show that dispersion forces become effective for near isothermal systems with sharp fin-groove corners and the film profile experiences an abrupt change, a \textit{slope break}, in the close proximity of the corner.  The current study is the first computational confirmation of this behavior in the literature.

\vspace{0.2cm}
\end{abstract}
\vspace{-0.6cm}
\begin{center} \noindent \HRule \\ \end{center}

\pagebreak

\section{Introduction}
\label{sec:intro}

The enhancements in the electronic chip production technologies have been made absolute improvements in the chip capacity, which lead to a huge increase in the chip power density. Proper operation of the chips with high power density is only possible with sufficient cooling. Thermal scientists have long been working on the efficient cooling techniques for the electronics in both device and chip levels. Phase-change cooling methods use the advantage of high latent heat of vaporization, which enable removing high amount of heat with very small temperature differentials \cite{faghri1995}. Furthermore, phase-change methods do not suffer from the high pumping power requirements encountered in single-phase cooling techniques.

Heat pipes, which are passive devices and use the phase-change heat transfer mechanism for heat spreading, have been widely used in various terrestrial, aviation and space applications since their establishment in 1964 \cite{grover1964}. In a heat pipe, which is a closed and sealed system, the working fluid evaporates at the evaporator section and the vapor flows into the condenser section due to the density gradient. The vapor condenses  at the condenser section and the condensate flows back to the evaporator by capillary pumping created by the virtue of the wick structures integrated on the inner wall of the heat pipe. In order to estimate the overall performance of a heat pipe, it is essential to model all these physical phenomena ensuing in the heat pipe. The wick structures are commonly in the form of sintered grains, wire meshes and micro-grooves in conventional applications. Among them, the micro-grooves are the mostly investigated structures for the thermal characterization and performance evaluation of heat pipes because of their advantages in developing numerical models and the relative ease of manufacturing \cite{hopkins1999,lips2010combined,alijani2018a,omur2018,alijani2018b}. Moreover, considering the chip-level applications, it is not feasible to apply sintered grains and wire meshes, while micro-grooves can be engraved on the semiconductors \cite{peterson1993,harris2010,kundu2015}. 

Modeling the physics inside the heat pipe is complicated, since it involves different scale problems: macro-scale in axial liquid and vapor flow and micro-scale in thin film evaporation and condensation near the groove edges. While the modeling of evaporation is widely studied \cite{lefevre2008,lips2010,do2008,do2010,odabasi2014,wayner1976,holm1979,moosman1980,mirzamoghadam1988,busse1992,dasgupta1994,wee2005,wang_garimella2007,ma2008,bertossi2009,narayanan2011,du2011,biswal2011,biswal2013,bai2013,ball2013,kou2015,akkus2016,akkus2017}, studies on the condensation modeling remain restricted \cite{kamotani1976,zhang2001,lefevre2008,lips2010,do2008,do2010,odabasi2014}. In the condenser section of a grooved heat pipe, the thickness of liquid on the fin tops is much smaller than the one inside the groove, which makes the resistance to heat transfer much lower on the fin tops. Therefore, majority of the condensation occurs on the fin tops. All of the condensate formed on the fin tops is assumed to flow into the groove in the existing condensation models \cite{kamotani1976,zhang2001,lefevre2008,lips2010,do2008,do2010,odabasi2014}, which reduces the condensation modeling to a two-dimensional problem in the condenser cross section by neglecting the possible axial flow on the fin tops. Upper boundary of this problem is the liquid-vapor interface, which is unknown \textit{a priori}. As a common simplification, this free surface's profile was assumed as a 4\textsuperscript{th} order polynomial in most of the studies \cite{kamotani1976,do2008,do2010,odabasi2014}, and yet there exist some studies which utilized the hydrodynamic models or CFD methods for the free surface estimation \cite{zhang2001,lefevre2008,lips2010}. Moreover, the slope of the free surface at the fin top-groove corner was assumed to be continuous by the previous studies \cite{kamotani1976,lefevre2008,lips2010,do2008,do2010,odabasi2014}. However, experimental study of Lips \textit{et al.} \cite{lips2010} reported a \textit{slope break} for the free surface at the fin top-groove corner, which made the assumption of continuous slope questionable. They indicated that their hydrodynamic model \cite{lefevre2008,lips2010} overestimates the film thickness on the fin top and concluded that, the possible effect of disjoining pressure or the cross-flow on the fin top, which were both neglected in their model, may be the underlying reason of this controversy.

Disjoining pressure is a surface phenomenon governed by molecular forces similar to the capillary pressure. Pressure jump across the liquid-vapor interface, \textit{i.e.} pressure difference between the liquid and vapor phases, is related to disjoining and capillary pressures by augmented Young-Laplace equation. Although the effect of disjoining pressure on the pressure jump was considered in the calculation of condensing mass flux in a few previous modeling attempts \cite{do2008,do2010}, the contribution of disjoining pressure gradient to the liquid flow along the condensing film was neglected by previous studies. The existing thin film condensation models, which were built on aforementioned assumptions, have not been experimentally validated yet, even for the simple geometries, \textit{i.e.} grooved heat pipes \cite{lips2016}. Therefore, validity of previous assumptions has not been justified by a comprehensive numerical model or an experimental observation. 

The current study aims to investigate the effect of disjoining pressure on the film thickness profile of the condensate on the fin top, and to scrutinize the validity of the continuous slope of the free surface assumption at the corner. For this purpose, a unidirectional flow solver is developed and the condensation problem on a fin-groove system is solved. The unidirectional flow solver is utilized in a solution domain, which starts at a point on the vertical wall of the groove and ends at the centerline of the fin; thereby, the fin-groove corner is kept inside the solution domain and the need for using a boundary condition at the corner is eliminated. In the algorithm developed, the conservation of mass and momentum equations, augmented Young-Laplace equation and the condensation mass flux equation based on the kinetic theory of gases (Kucherov-Rikenglaz equation) are solved simultaneously to calculate the film thickness profile on the fin top. The novel solution methodology utilized in the present study enables the inclusion of the thinnest part of the liquid film near the groove edge to the condensation model. Therefore, this work takes the first step towards a comprehensive understanding of molecular forces on the condensing liquid film profile formed in a fin-groove system. 

The article is organized as follows: in Section~\ref{sec:modeling}, the problem domain is specified and the lubrication approximation for the flow is elucidated. The governing equations are presented together with the solution methodology. In Section~\ref{sec:results}, the film thickness profiles for different cases are reported and the effect of superheat, disjoining pressure and corner radius on the film profiles are investigated. The conclusions are presented in Section~\ref{sec:conclusion}. 

\section{Modeling}
\label{sec:modeling}

The current study considers the condensation problem in a fin-groove system composed of infinite number of grooves in lateral direction. Vapor phase of the working fluid condenses to its liquid phase on the solid surface, which is sub-cooled at a constant temperature. The length of each groove is also sufficiently long and the condensed liquid is assumed to be discharged from the groove bottom at a flow rate equal to the condensation flow rate such that the variations in axial direction are negligible, which makes the problem two-dimensional in the cross sectional plane of the fin-groove system. The height of the liquid inside the groove may differ depending on the initial liquid amount in the groove and the discharge flow rate. Thus, there exists different steady state solutions of the problem corresponding to different liquid heights inside the channel.

\subsection{Physical domain} 

\begin{figure}
\includegraphics[scale=0.9]{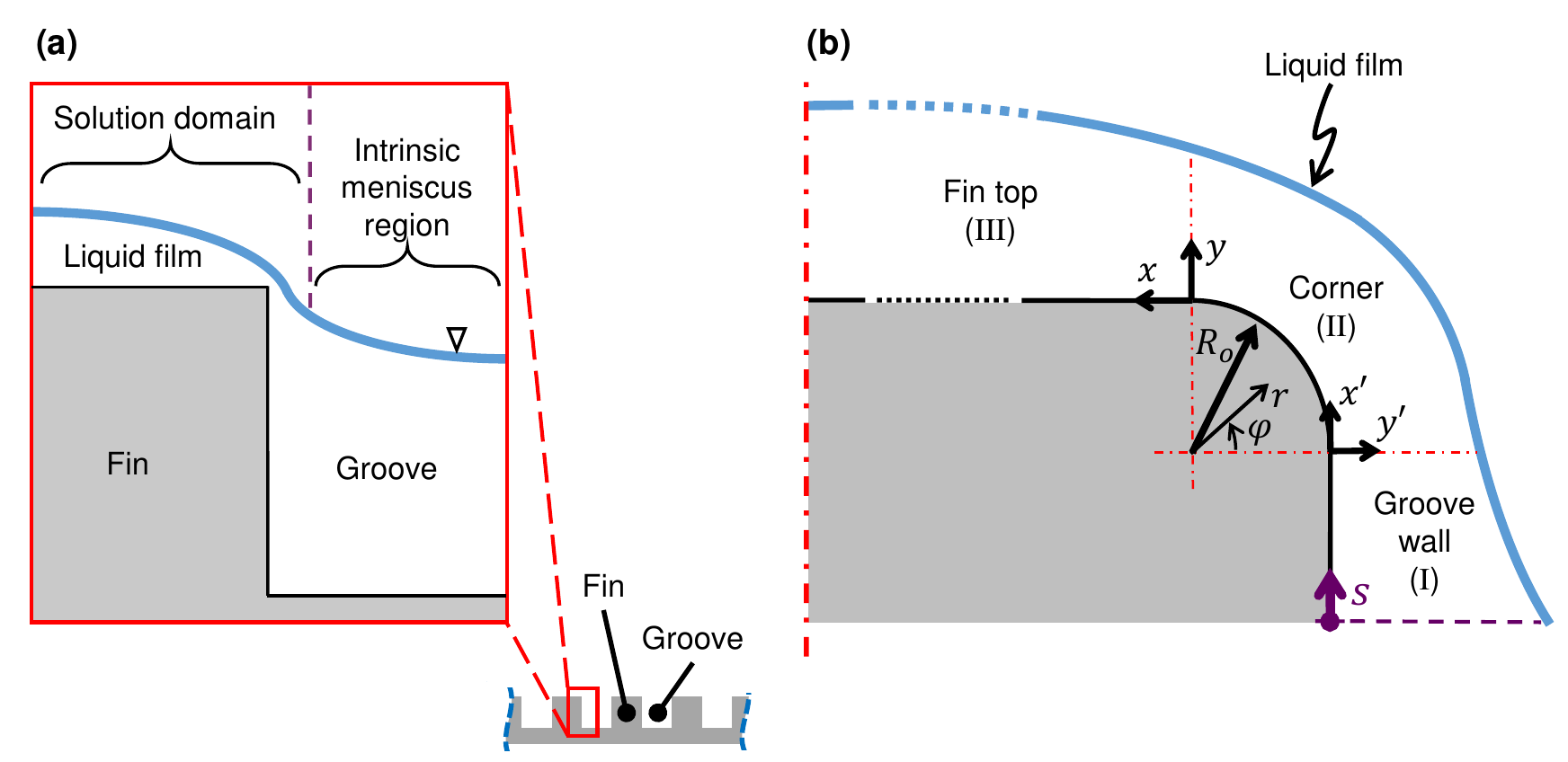}
\centering
\caption{Physical domain for the problem. (a) \textit{Problem domain} is defined between the center planes of an adjacent fin and groove pair. (b) \textit{Solution domain} includes the fin top and the close proximity of fin-groove corner extending to the groove.}
\label{problem_domain}
\end{figure}

The center planes of an adjacent fin and groove pair are the symmetry planes due to the repeating pattern of the fin-groove system. \textit{Problem domain} is defined between these symmetry planes as shown in Fig.~\ref{problem_domain}a. The liquid-vapor interface within the problem domain is practically divided into two parts: the intrinsic (bulk) meniscus region and the thin film region on the fin top. While the former one is associated with relatively low condensation rates and a near circular profile, the profile of the latter one is of interest due to strong condensation, especially near the corner region. Therefore, the current study strives to solve the film profile on the fin top only. \textit{Solution domain} for this film starts at a point on the groove wall and ends at the symmetry line of the fin and it is composed of three regions as shown in Fig.~\ref{problem_domain}b: groove wall (I), corner (II), and fin top (III). The fin-groove corner is approximated by a cylindrical surface with radius, \textit{$R_o$}. The governing equations are formulated using Cartesian coordinates for planar surfaces and polar coordinates for the cylindrical surface. Origins of the Cartesian~($x$,$y$) and polar~($r$,$\varphi$) coordinate systems are shown in Fig.~\ref{problem_domain}b. During the solution, the governing equations are transformed into the surface coordinate, \textit{s}, the origin of which is also displayed in Fig.~\ref{problem_domain}b. For the planar surfaces, where surface coordinate is linear, the differentials are the same for $x$- and \textit{s}-coordinates, which makes the transformation from Cartesian to the surface coordinate straightforward. However, switching to surface coordinate along the cylindrical surface requires the following transformation for the differentials: $ds=R_od\varphi \ $.

\subsection{Lubrication assumption} 

Creeping flow assumption is applied, which is similar to the approximations made in case of the flow of a lubricant inside a journal bearing. The main simplicity that the lubrication assumption brings is the use of a parabolic velocity profile on the cylindrical surface of the journal bearing. The lubrication assumption is valid as long as the diffusion time scale, $t_{di\mathit{ff}}$, is sufficiently smaller than the convection time scale, $t_{conv}$, for the liquid flow. The condition for the lubrication assumption can be written as the functions of ratio of film thickness, $\delta$, to  the extend of flow, $L_{flow}$, and Reynolds number:

\begin{equation}
\frac{t_{di\mathit{ff}}}{t_{conv}} \sim \mathcal{O}  \Big( \frac{\delta}{L_{flow}}\mathrm{Re} \Big) \ll 1 .
\label{lub-pl}
\end{equation}

\noindent For the above condition, Reynolds number is defined as $\mathrm{Re}= u_x\delta \big/\nu$ on a planar surface, where $u_x$ is the mean velocity parallel to planar surface and $\nu$ is the kinematic viscosity. If the length of the surface in the direction of the flow is much longer than the film thickness or Reynolds number is small, lubrication assumption can be utilized for the liquid flow on the planar surface. For the flow of condensate on the fin top \cite{kamotani1976,lefevre2008,lips2010,do2008,do2010,odabasi2014} or the flow of liquid on a heated planar substrate towards the contact line \cite{lefevre2008,lips2010,do2008,do2010,odabasi2014,wayner1976,holm1979,moosman1980,mirzamoghadam1988,busse1992,dasgupta1994,wee2005,wang_garimella2007,ma2008,bertossi2009,narayanan2011,du2011,biswal2011,biswal2013,bai2013,ball2013,kou2015,akkus2016,akkus2017}, 
 lubrication assumption has been widely utilized. For a cylindrical surface, on the other hand, Reynolds number is defined as $\mathrm{Re} = u_\varphi\delta\big/\nu$, where $u_\varphi$ is the mean angular velocity in $\varphi$-direction, and the extent of the flow is expressed as the function of the radius of the cylindrical surface ($L_{flow}=R_o\varphi$). For the journal bearing problems, the radial clearance between the journal and the bearing is extremely small ($\delta/R_o\ll1$), therefore, the lubrication assumption holds even for high rotational speed of the shaft. In the condensate flow over fin-groove corner, on the other hand, Reynolds number is excessively small, due to the very low velocities. Therefore, the lubrication assumption holds even for comparable magnitudes of the film thickness and corner radius.
 
\subsection{Flow and condensation model} 

The mass balance within the condensing film can be expressed in terms of the mass flow rate per unit depth along the surface coordinate, $\dot m'$, and the condensation mass flux at the free surface, $\dot m_c^{''}$, as follows: 

\begin{equation} 
\frac{{d\dot m'}}{{ds}} =  - \dot m_c^{''}.
\label{mass}
\end{equation}

\noindent Utilizing the lubrication assumption, \textit{i.e.} neglecting inertial and longitudinal diffusive terms, conservation of linear momentum reduces to Eqs.~(\ref{mom-pl})~and~(\ref{mom-cyl}) for planar and cylindrical surfaces, respectively.

\begin{subequations}
\begin{equation}
\frac{dp_l}{ds}=\mu\frac{d^2u}{dy^2},
\label{mom-pl}
\end{equation}
\begin{equation}
\frac{dp_l}{ds}=\mu\frac{d^2u}{dr^2},
\label{mom-cyl}
\end{equation}
\end{subequations}

\noindent where, $p_l$ is the liquid pressure, $\mu$ is the dynamic viscosity and $u$ is the velocity in the $s$-direction.  Eq.~(\ref{mom-pl}), expressed for the fin top, is also applicable on the groove wall, with replacing $x$ by $x'$  and $y$ by $y'$.  In order to obtain the velocity profiles, Eqs.~(\ref{mom-pl})~and (\ref{mom-cyl}) are integrated from the wall surface to the liquid-vapor interface invoking no-slip boundary condition at the wall and zero-shear boundary condition at the interface. Using these velocity profiles, the mass flow rates (per unit depth) along the surface coordinate are obtained as,

\begin{subequations}
\begin{equation}
\dot m'=-\frac{1}{3\nu }\frac{dp_l}{ds}\,{\delta}^3,               
\label{m_pl}
\end{equation}
\begin{equation}
\dot m'=\frac{1}{\nu }\frac{dp_l}{ds}\,\Bigg[\frac{(R_o+\delta)^3-R_o^3}{6}-\frac{(R_o+\delta)((R_o+\delta)^2-{R_o}^2)}{2}+ \frac{2\delta^2 R_o+\delta {R_o}^2}{2}\Bigg],               
\label{m_cyl}
\end{equation}
\label{m_dot}
\end{subequations}

\noindent for planar and cylindrical surfaces, respectively. The liquid pressure, $p_l$, is calculated using the well--known augmented Young-Laplace equation,
\begin{equation}
p_v-p_l=p_c+p_d,
\label{ayle}
\end{equation}
\noindent where, capillary pressure, $p_c$, is defined as,

\begin{subequations}
\begin{equation}
p_c=\sigma\frac{\delta_{ss}}{{\left( {1 + {{\delta_s}^2}} \right)}^{3/2}},
\label{pc_pl}
\end{equation}

\begin{equation}
p_c=\sigma\frac{\left( \delta+R_o \right){R_o}^2\delta_{ss}-2{R_o}^2\delta_s^2-\left( \delta+R_o \right)^2}{{\left( {\left( \delta+R_o \right)^2 + {{R_o}^2{\delta_s}^2}} \right)}^{3/2}},
\label{pc_pl}
\end{equation}
\end{subequations}

\noindent for planar and cylindrical surfaces, respectively. In the absence of retardation and structural effects, disjoining pressure, $p_d$, of a non-polar liquid film can be expressed by the power relation \cite{derjaguin1957},

\begin{equation}
p_d=\frac{A_d}{\delta^3}.             
\label{pd}
\end{equation}

\noindent Many studies in the literature, on the other hand, used the power relation for even strong polar liquids \cite{do2008,do2010,wayner1976,ma2008}, although it is not valid due to the presence of short range intermolecular forces (hydrogen bonding, hydration forces etc.) in addition to the long range intermolecular forces (Lifshitz-van der Waal forces). In the present study, a non-polar liquid (octane) is used to refrain from an improper use of the power relation. Assuming constant vapor pressure, the liquid pressure gradient can be written as,

\begin{equation}
\frac{dp_l}{ds}=\frac{d}{ds}\left( p_c+p_d \right).               
\label{dpds}
\end{equation}

\noindent Mass flux of the phase-change at the interface is calculated based on Kucherov-Rikenglaz equation~\cite{kucherov1960} and it can be expressed as functions of the temperature difference between the interface and vapor (superheat) and the pressure difference across the interface (pressure jump) \cite{wayner1971}. The interface temperature can be replaced by the wall temperature assuming pure conduction within the liquid film \cite{moosman1980}: 

\begin{subequations}
\begin{equation}
\dot m_c^{''} = \frac{{a\left( {{T_w} - {T_v}} \right) - b\left( {{p_v} - {p_l}} \right)}}{{1 + 
a\delta h_{lv}/k_l}},
\label{m7}
\end{equation}

\begin{equation}
a= \frac{2c}{2-c} \Bigl(\frac{M}{2\pi R_u T_{lv} } \Bigr)^{1/2}\frac{p_v Mh_{lv}}{ R_u T_v T_{lv}},
\label{m8}
\end{equation}
\begin{equation} 
b= \frac{2c}{2-c} \Bigl(\frac{M}{2\pi R_u T_{lv }} \Bigr)^{1/2}\frac{p_v V_l}{ R_u T_{lv}}, 
\label{m9}
\end{equation}
\end{subequations}
where, $T_w$, $T_v$, $T_{lv}$ are the wall, vapor and liquid-vapor interfacial temperatures, respectively; $p_v$ is the vapor pressure; $h_{lv}$ is the latent heat; $k_l$ is the thermal conductivity of liquid; $M$ is the molar mass; $V_l$ is the molar volume; $R_u$ is the universal gas constant; and $c$ is the accommodation coefficient, which is taken as unity~\cite{do2008,dasgupta1994,du2011,bai2013,kou2015}.

\begin{figure} [b!]
\includegraphics[scale=.35]{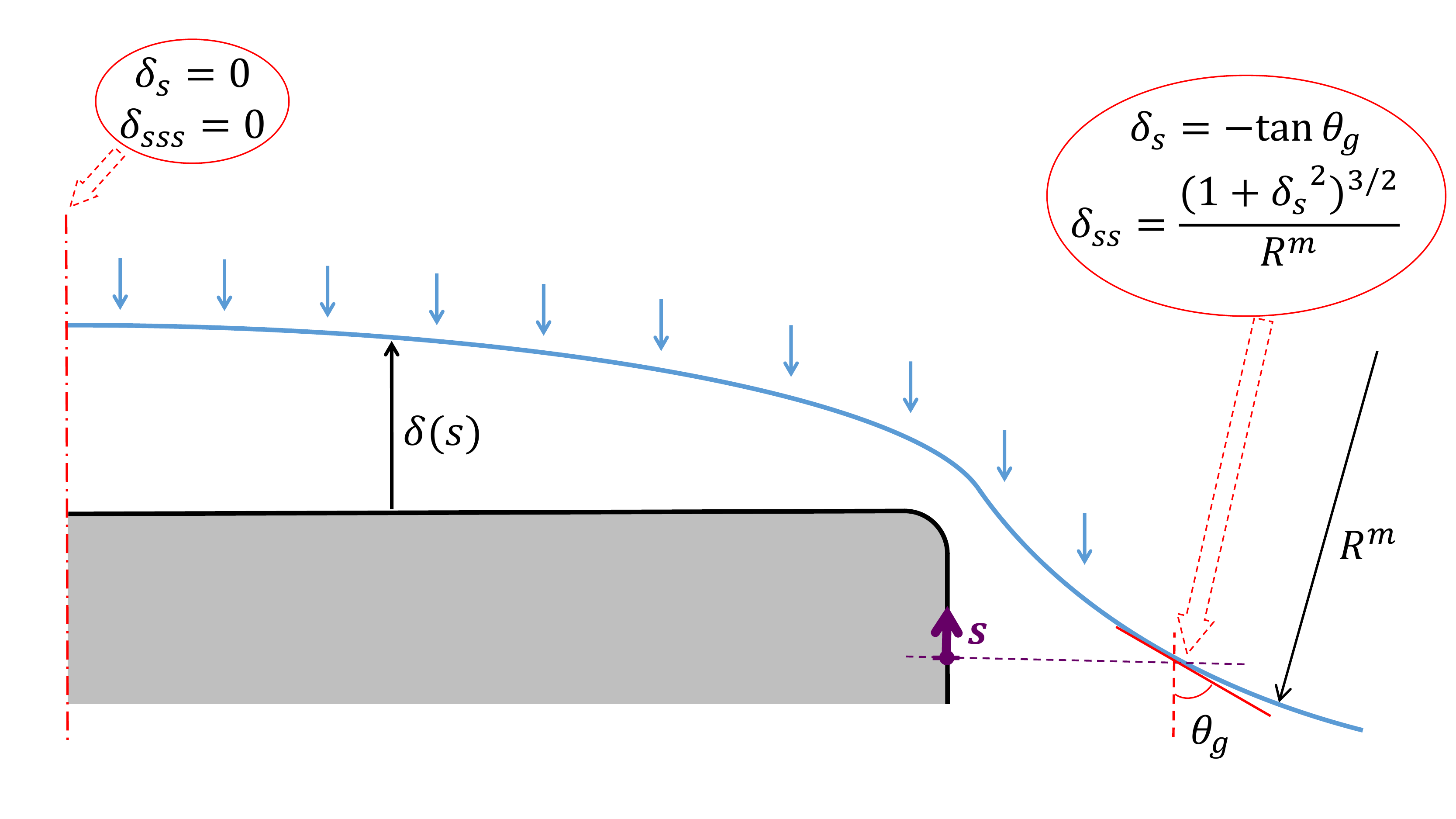}
\centering
\caption{Boundary conditions}
\label{domain_bc}
\end{figure}

Substitution of the mass flow rate per unit depth, $\dot m'$, and the condensation mass flux at the free surface, $\dot m_c^{''}$, into mass balance equation, Eq.~(\ref{mass}), yields the following relations,
\begin{subequations}
\begin{equation} 
- \frac{1}{{3\nu }}\frac{d}{ds}\Big( \delta^3\frac{dp_l}{ds} \Big) =  - \frac{{a\left( {{T_w} - {T_v}} \right) - b\left( {{p_v} - {p_l}} \right)}}{{1 + a\delta h_{lv}/k_l}},
\label{ode_pl}
\end{equation}
\begin{multline}
\frac{1}{\nu }\frac{d}{ds}\Bigg[\frac{dp_l}{ds}\,\Bigg(\frac{(R_o+\delta)^3-R_o^3}{6}-\frac{(R_o+\delta)((R_o+\delta)^2-{R_o}^2)}{2}+ \\ \frac{2\delta^2 R_o+\delta {R_o}^2}{2}\Bigg)\Bigg] = - \frac{{a\left( {{T_w} - {T_v}} \right) - b\left( {{p_v} - {p_l}} \right)}}{{1 + a\delta h_{lv}/k_l}}, 
\label{ode_cyl}
\end{multline}
\end{subequations}
for planar and cylindrical surfaces, respectively. Eq.~(\ref{ode_pl})~and (\ref{ode_cyl}), 4\textsuperscript{th} order ODEs of film thickness, require four boundary conditions, which are presented in Fig.~\ref{domain_bc} for the current modeling approach.  The solution starts at a point on the groove wall, where the radius of curvature of the meniscus, $R^m$, and the edge angle of the liquid-vapor interface inside the groove, $\theta_{g}$---the minimum value of which is the apparent contact angle of liquid on the substrate---, are known. The first and second derivatives of the film thickness at the boundary of the problem located in the groove wall region are calculated based on $R^m$ and $\theta_{g}$. The other two boundary conditions, the first and third derivatives of the film thickness, are defined at the boundary of the problem located at the central plane of the fin, based on the symmetry condition.

\subsection{Solution approach} 

\begin{figure}
\centering
\includegraphics[width=100mm]{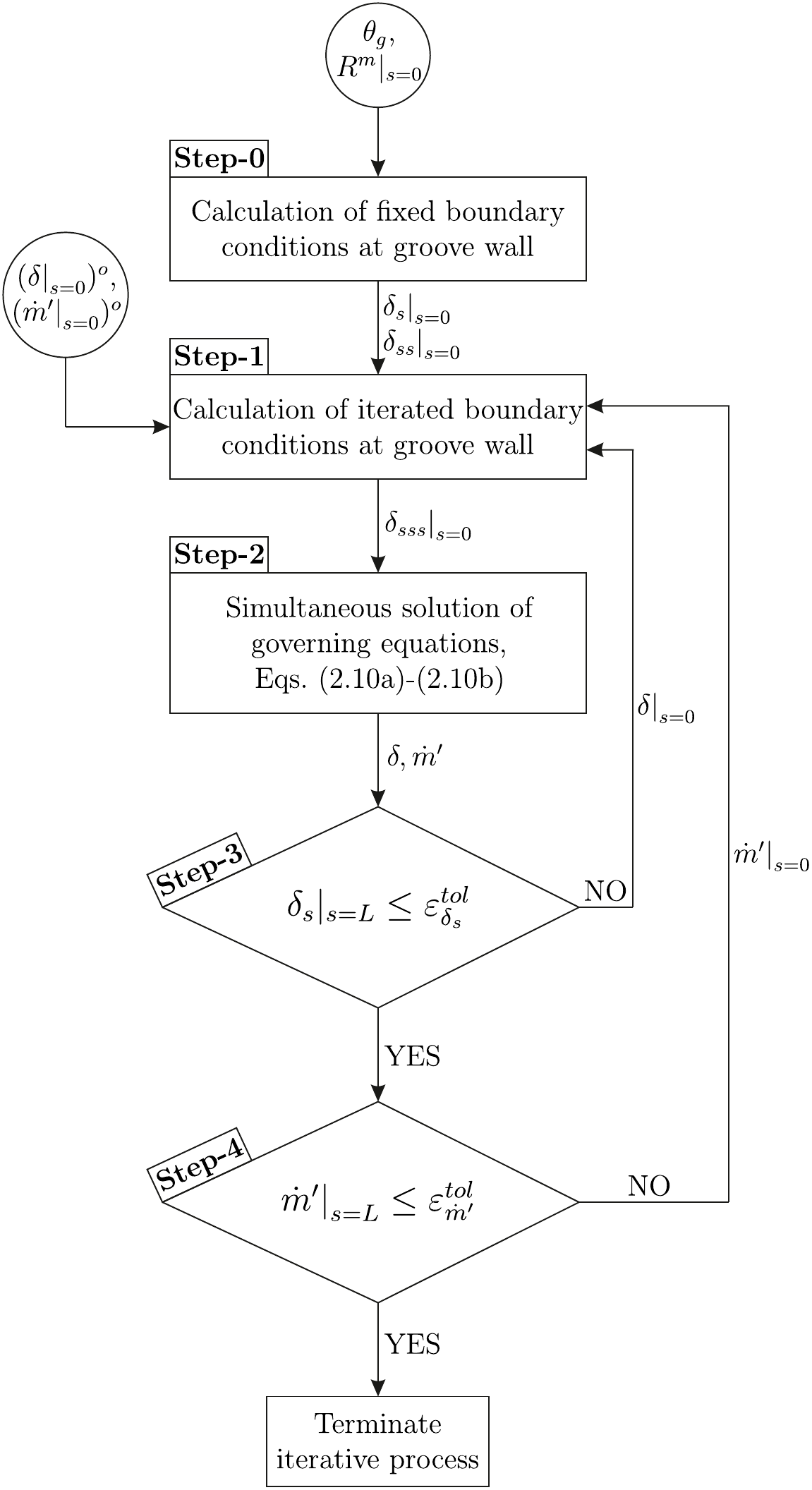}
\caption{Iterative computational scheme}
\label{flow}
\end{figure}

In the described problem, the distribution of the condensing film profile is unknown \textit{a priori}. The objective of the solver developed is to calculate the film thickness along the surface coordinate in accordance with the boundary conditions. The solution starts at a position on the groove wall with two specified boundary conditions at $s=0$ (Fig.~\ref{domain_bc}) and the initial guesses for the film thickness and mass flow rate. Thus, using four boundary conditions at $s=0$, the solver calculates the film thickness distribution up to  the central plane of the fin, where two symmetry boundary conditions are expected to hold. In order to obtain the desired boundary conditions at the end of the solution domain, initial guesses for the starting film thickness and mass flow rate are iterated using two nested secant iteration loops. The flowchart for the solution algorithm is given in Fig.~\ref{flow} and the details of the solution steps are explained below.

\textit{Step-0.} This step yields two boundary conditions, which are not subjected to iteration during the solution algorithm, based on two geometrical inputs defined at the starting point of the problem. More specifically, the first and second derivatives of film profile at $s=0$ are calculated using the edge angle and the radius of curvature of the liquid film inside the groove, as described in Fig.~\ref{domain_bc}.  

\textit{Step-1.} This step aims to calculate the other two boundary conditions, which are the film thickness and the third derivative thereof, to initiate the solution process at the starting mathematical boundary. While the film thickness is directly provided as an input, the mass flow rate of the condensate, which is the functions of film thickness and its first, second and third derivatives (Eqs.~(\ref{m_dot})--(\ref{dpds})), is utilized to calculate the third derivative at $s=0$. The inputs provided in this step are iterated during the solution in order to match the target boundary conditions at the end of the domain.

\textit{Step-2.} This step calculates the film profile (and the distribution of mass flow rate) along the problem domain by solving the governing equations (Eqs.~(\ref{ode_pl})~and~(\ref{ode_cyl})) based on the boundary conditions estimated in the previous steps. In the solution procedure, the solution domain is discretized into successive strips and the mass balance is secured in each strip as follows:

\begin{equation} 
\frac{{\dot m'}_{i+1}-{\dot m'}_i}{\Delta s} =  - {\dot m_{c(i,i+1)}^{''}},
\label{mass_disc}
\end{equation}
where, $i$ and $(i+1)$ denotes the consecutive edges of a strip and $\dot m_{c(i,i+1)}$ is the average of the condensation mass fluxes at $i^{th}$ and $(i+1)^{th}$ points. 

\textit{Step-3.} This step checks if the zero slope boundary condition at the symmetry line ($s=L$) holds, using a sufficiently small tolerance value, $\varepsilon^{tol}_{\delta_s}$. If the slope is not sufficiently small, the previous step is repeated with a new film thickness guess at $s=0$, else the \textit{Step-4} starts.

\textit{Step-4.} The symmetry boundary condition specified for the third derivative of film thickness given in Fig.~\ref{domain_bc} implies that the mass flow rate at the at the symmetry line of the fin is zero. This step checks if the mass flow rate is less than a sufficiently small tolerance value, $\varepsilon^{tol}_{\dot m'}$. If the flow rate is higher than the tolerance, the \textit{Step-1} is repeated with a new mass flow rate guess at $s=0$, else the the iterative process is terminated.

To summarize, the algorithm iteratively seeks for  film thickness and mass flow rate values at $s=0$, which render the slope of the film profile and the mass flow rate zero on the symmetry line, utilizing two nested secant loops for root finding. One of the main complexities in this solution procedure arises at the locations of transitions from planar groove wall to cylindrical corner and from cylindrical corner to planar fin top wall (at $\varphi=0$ and $\varphi=\pi/2$, respectively). The film thickness and its first, second and third derivatives with respect to surface coordinate, $s$, are continuous within the planar and cylindrical wall regions. However, the solid surface curvature and the derivatives of the solid surface normal vector are discontinuous at the transition points specified. Accordingly, the derivatives of the film thickness with respect to surface coordinate are also discontinuous at these points. A set of matching conditions are required to match the derivatives at the transition points between the planar and cylindrical wall regions~\cite{kamotani1976}. Therefore, matching conditions at the transition points were written by implementing  the continuity of the film thickness, liquid pressure and mass flow rate together with the smoothness of the free surface profile in the numerical integration process. 

\section{Results and discussion}
\label{sec:results}

Liquid film profiles on the fin top are obtained for various cases and the effect of disjoining pressure on the film profile is investigated by focusing on the region near the fin top-groove corner. The working fluid is octane and the solid is silicon similar to previous phase-change studies \cite{schonberg1992, wang_garimella2007}. The temperature and the pressure of the vapor phase, together with the latent heat of vaporization are selected following \cite{schonberg1992}. The numerical value of $3.18\times10^{-21}\unit{J}$ for the dispersion constant was used, similar to \cite{schonberg1992, wang_garimella2007}.  Other thermophysical properties of the liquid and vapor are evaluated at 343$\unit{K}$ using NIST Chemistry WebBook \citep{linstrom2001}. Thermophysical properties and geometrical parameters used in the model are summarized in Table~\ref{props}.

\begin{table}[h]
\caption{Thermophysical properties and geometrical parameters used in the model}

\begin{center}
\begin{tabular}{|l|l|l|}
\hline 
Vapor temperature (K)                     & $ T_v $       & 343 \\
\hline  
Vapor pressure (Pa)                       & $ p_v $       &  15869  \\ 
\hline
Density of saturated vapor (kg$\unit{m^{-3}}$)      & $ \rho_v $      & 0.63964 \\    
\hline
Density of liquid (kg$\unit{m^{-3}}$)      & $ \rho_l $      & 661.38 \\ 
\hline 
Latent heat (J$\unit{kg^{-1}}$)         & $ h_{lv} $    & $ 3.398\times 10^{5} $ \\ 
\hline 
Surface tension (N$\unit{m^{-1}}$)                      & $\sigma$      & 0.016953 \\ 
\hline
Dynamic viscosity of liquid ($\unit{Pa}\unit{s}$)    & $ \mu _l $       & $ 3.1929\times 10^{-4} $ \\ 
\hline
Thermal conductivity (W$\unit{m^{-1}} \unit{K^{-1}}$)    & $ k_l $       & 0.11136 \\ 
\hline 
Molar mass of liquid (kg$\unit{mol^{-1}}$)              & $ M $         & 0.11423 \\ 
\hline 
Molar volume of liquid ($ \rm m^{3} \unit{mol^{-1}}$) & $ V_l $       & $1.7271\times10^{-4}$ \\ 
\hline 
Accommodation coefficient                  & $c$ & 1 \\ 
\hline 
Dispersion constant (J)                    &  $ A_d $     & $ 3.18\times 10^{-21} $ \\ 
\hline
Radius of meniscus in groove ($\mu$m)                       & $ R^m $        & $ 800 $ \\ 
\hline 
Fin top length ($\mu$m)                       & $ L_{fin} $        & $ 50 $ \\ 
\hline 
\end{tabular} 
 
\end{center}
\label{props}
\end{table} 

\subsection{Effect of superheat} 

The condensation mass flux depends on the film thickness, pressure jump across the liquid-vapor interface, the temperature difference (superheat) between the wall and vapor, and the thermophysical properties of the fluid. The film thickness and pressure jump distributions are the results of the solver. Superheat, on the other hand, is an input and directly affects the mass flux. Therefore, the effect of superheat on the results is of primary interest. While the corner radius, $R_o$, is taken as $30\unit{nm}$, the values used for superheat range between $1\unit{K}$ and  $10^{-3}\unit{K}$, which is nearly isothermal. The edge angle inside the groove, $\theta_{g}$, is 30$\degree$. 

\begin{figure} [t]
\centering
\includegraphics[width=90mm]{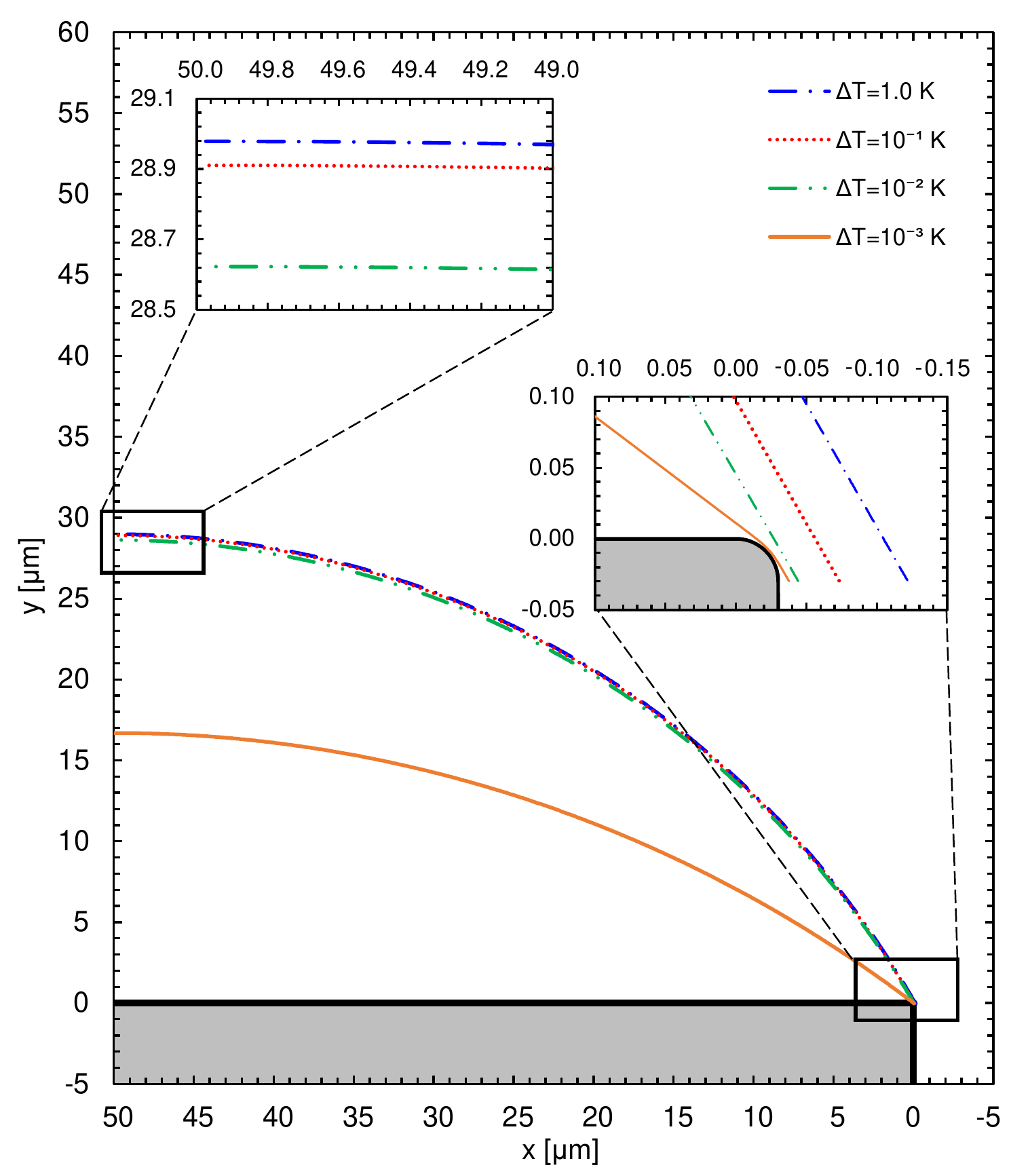}
\caption{Effect of superheat on the film thickness profile on the fin top}
\label{superheat1}
\end{figure}

The film thickness profiles obtained for four different superheat values are presented in Fig.~\ref{superheat1}. There is a slight difference in the film thickness profiles on the fin top for 1.0, $10^{-1}$, and $10^{-2}\unit{K}$ superheat. However, for $10^{-3}\unit{K}$, there is a significant decrease in the film thickness. When the insets showing the variations of film thickness profiles near the symmetry line and corner region are examined, the effect of superheat on the film profiles is apparent: the film thickness decrease with decreasing temperature difference. The slope of the film profiles, on the other hand, do not change significantly from the groove side to the fin top for superheat values between $10^{-2}$ and  $1.0\unit{K}$. However, for $10^{-3}\unit{K}$ superheat, the film profile conforms to the solid substrate surface at the corner region, and accordingly, at the fin top starting point ($x=0$), the slope of the film profile is smaller than the slopes obtained for higher superheat values, which results in a thinner film on the fin top. Therefore, the continuous slope assumption at the fin top-groove corner, widely used in the literature, is not valid for the near isothermal cases of the problem investigated.

\begin{figure} [h]
\centering
\includegraphics[scale=0.7]{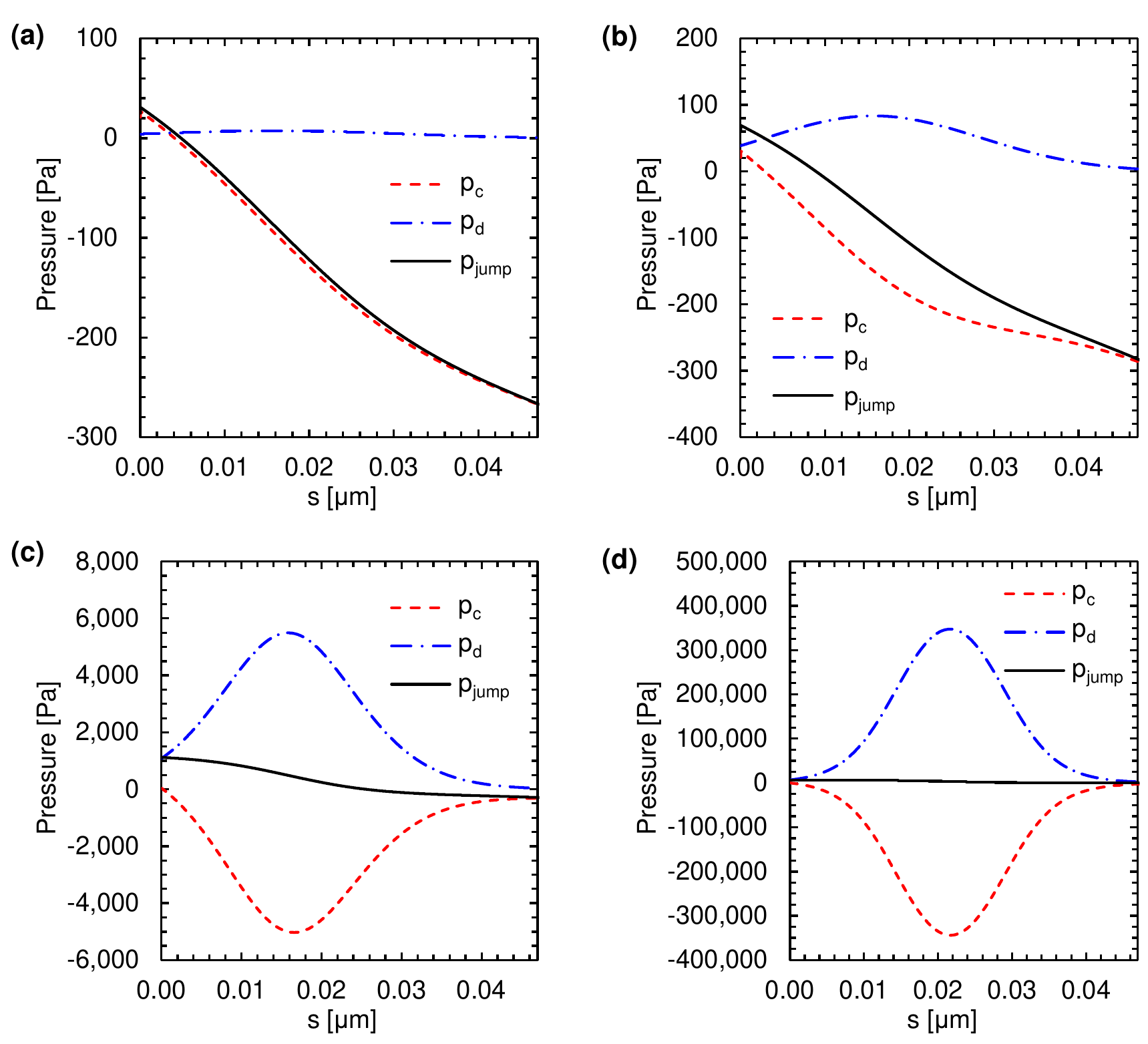}
\caption{Capillary pressure, disjoining pressure and pressure jump at corner region for superheat values of (a) $1.0\unit{K}$, (b) $10^{-1}\unit{K}$, (c) $10^{-2}\unit{K}$, (d) $10^{-3}\unit{K}$}
\label{pc-pd-pj}
\end{figure}

For $10^{-3}\unit{K}$ superheat, the free surface makes a sharper turn at the corner region. Therefore, there is a substantial change in the free surface curvature in this region, which indicates the existence of a considerable capillary pressure gradient. Moreover, the film thickness at the corner is much lower for $10^{-3}\unit{K}$ superheat, which means the effect of the disjoining pressure is also higher. The capillary pressure, disjoining pressure and their sum (pressure jump) are plotted in Fig.~\ref{pc-pd-pj} for all superheat values. For the largest superheat, film thickness profile is the thickest due to the elevated condensation rates. At this superheat, the effect of disjoining pressure is negligible (see Fig.~\ref{pc-pd-pj}a). However, as the superheat decreases, the effect of the disjoining pressure becomes more pronounced. For the smallest superheat, disjoining pressure reaches its highest magnitudes due to the thinner film profile. This large increase in disjoining pressure is compensated by the capillary pressure as shown in Fig.~\ref{pc-pd-pj}d. Therefore, the gradient of the capillary pressure is also high at the corner region. The curvature of the film profile changes abruptly to create this high capillary pressure gradient, which leads to a sharp turn of the free surface in the corner region. The result is a film profile tracing the cylindrical wall surface for the smallest superheat. The symmetric distribution of disjoining pressure in Fig.~\ref{pc-pd-pj}d is also related with the conformal profile of the film due to the fact that the thinnest part of the film profile occurs near the mid-point of the cylindrical corner. For higher superheat values, the film profiles become thinnest, \textit{i.e.} peak disjoining pressure occurs, before the mid-point of the corner region (see Fig.~\ref{pc-pd-pj}b-c), since the effect of the disjoining pressure is small.

\subsection{Effect of disjoining pressure}

\begin{figure} [h]
\centering
\includegraphics[scale=0.8]{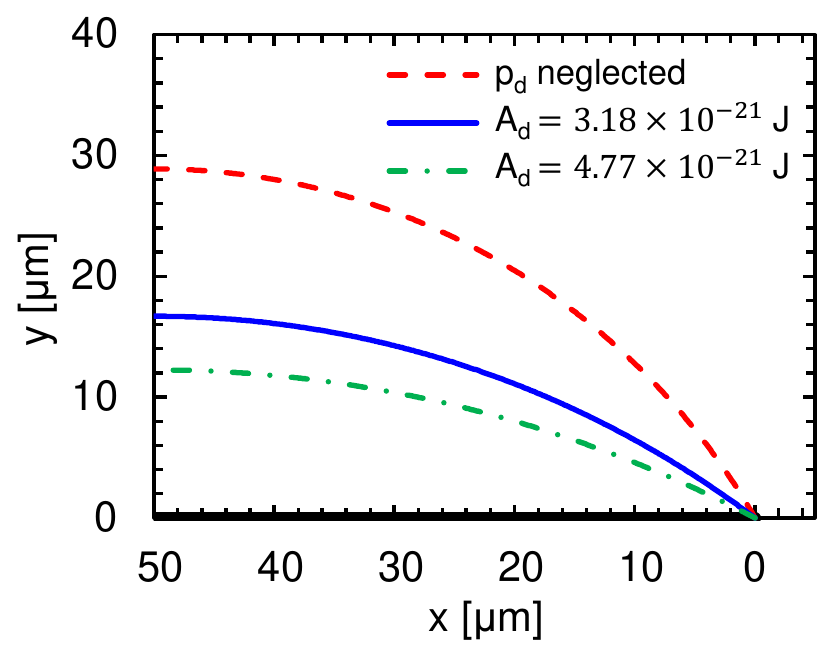}
\caption{Effect of dispersion constant on the film thickness profile ($\Delta T=10^{-3}\unit{K}$, $R_0=30\unit{nm}$)}
\label{hamakerEff}
\end{figure}

To investigate the effect of disjoining pressure on the film thickness, two additional cases are solved using $10^{-3}\unit{K}$ superheat: in the first one, the disjoining pressure is neglected; and in the second one, a  dispersion constant 1.5 times higher than the actual value ($4.77\times10^{-21}\unit{J}$) is used. The film thickness profiles obtained are presented in Fig.~\ref{hamakerEff}. When the disjoining pressure is neglected, the film thickness profile is very close to the one obtained for the $1.0\unit{K}$ superheat case, where the disjoining pressure effect is negligible due to the thicker film. While the total condensation mass flow rate (per unit depth) is $2.6\times10^{-9}\unit{kg}\unit{m^{-1}}\unit{s^{-1}}$ for the solution with $3.18\times10^{-21}\unit{J}$ dispersion constant, it is $1.4\times10^{-9}\unit{kg}\unit{m^{-1}}\unit{s^{-1}}$ in the absence of disjoining pressure suggesting a 46\% deficit in the condensing vapor rate.

The dispersion constant, $A_d$, is defined as, $A_d = A/6\pi$, where $A$ is the Hamaker constant. The Hamaker constant is obtained experimentally and it is not reported for many solid-liquid-gas systems in literature. In addition, there are some theoretical methods such as combining rule to calculate the Hamaker constants for the multi-material systems using the properties of individual materials, but they do not always result in a valid approximation~\cite{israelachvili2011}. Consequently, there is not a consensus about the Hamaker constants of many systems and the reported Hamaker constants lie in a wide range. In the current study, in addition to using a zero Hamaker constant (i.e. effectively eliminating the dispersion effect), a second value, 1.5 times the original is used, to asses the effect of the magnitude of disjoining pressure. This upper bound of the dispersion constant elevates the total condensation mass flow rate to $3.6\times10^{-9}\unit{kg}\unit{m^{-1}}\unit{s^{-1}}$ resulting in a 38\% excess condensation rate, since the higher disjoining pressure makes the free surface more conformal to the solid substrate at the corner region, which leads to a thinner film on the fin top. 

The results obtained for these three cases reveal that exclusion of the disjoining pressure effect in the condensation modeling results in misleading film thickness profiles and mass flow rates for the small superheat values in fin-groove systems. Moreover, film thickness profiles and flow rates are sensitive to the magnitude of the Hamaker constant. Therefore, depending on the type of molecular interactions of different solid and fluid materials, the system may be influenced by the dispersion forces even for higher superheats.

\subsection{Effect of corner radius}

Lubrication approximation is widely applied to model the liquid flow within thin films including the problems with phase-changing interface. In the current problem, thin film is positioned on both the fin top and groove wall surfaces, where the unidirectional flow of the condensate can be modeled using lubrication approximation along the solid surfaces. However, the condensate flowing into the groove changes its direction at the intersection of the fin top and groove wall (corner region), where the unidirectionality of the liquid flow is disturbed. In order to utilize the lubrication approximation throughout the solution domain, the corner was modeled as a cylindrical surface rather than a sharp, discontinuous edge. In this section, the effect of size of the corner radius on the film thickness profile is investigated. The results are presented for the near isothermal case ($10^{-3}\unit{K}$ superheat), since the difference between the film profiles is more prominent for this case.

\begin{figure} [h]
\centering
\includegraphics[scale=0.8]{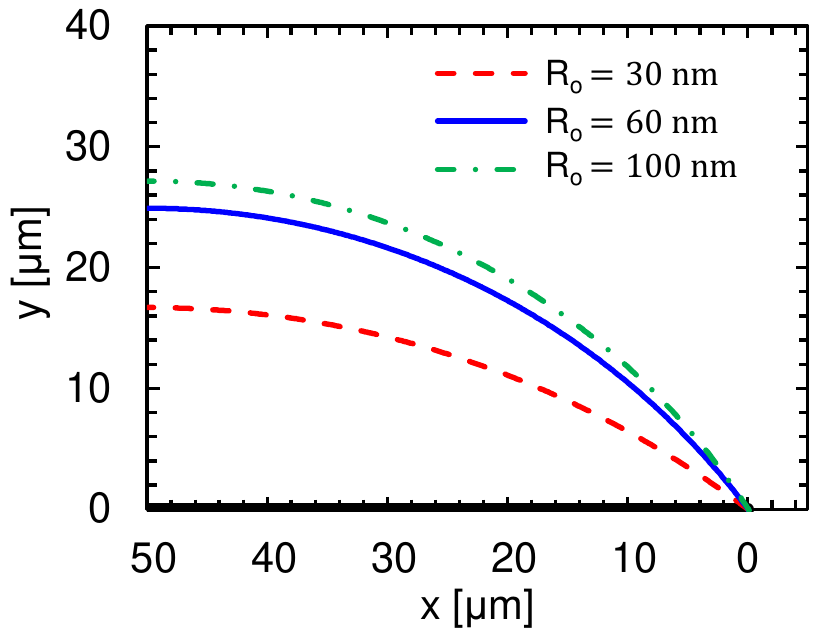}
\caption{Effect of corner radius on the film thickness profile ($\Delta T=10^{-3}\unit{K}$)}
\label{Ro}
\end{figure}

The film profiles for $30$, $60$, and $100\unit{nm}$ corner radius values are presented in Fig.~\ref{Ro}. As the corner becomes sharper, the free surface makes a sharper turn at the corner and the liquid film on the fin top is thinner. Considering this trend, it can be deduced that decreasing the radius further may lead to an even thinner film on the fin top. Thus, the effect of disjoining pressure may be significant even for higher superheat values when the corner is sharper. However, decreasing the corner radius further leads to the violation of the continuum and the lubrication assumptions of the current model. Therefore, the results for smaller corner radius values are not presented.

\begin{table} [t]
\caption{Minimum film thicknesses and time scale ratios at the corner region}
\begin{center}
  \begin{tabular}{c c c c}
    \hline
    \noalign{\vskip 3pt}
    \shortstack{Corner radius\\(nm)} & \shortstack{Superheat\\(K)} & \shortstack{Minimum film\\thickness (nm)} & \shortstack{Average time scale ratio\\ ($t_{di\mathit{ff}}/t_{conv}$)}  \\  \noalign{\vskip 3pt}
    \hline
\noalign{\vskip 2pt} 
    $30$ & $1.0$     & $75.7$ & $2.3\times10^{-2}$  \\ 
    $30$ &$10^{-1}$  & $33.6$ & $1.2\times10^{-3}$   \\ 
    $30$ &$10^{-2}$  & $8.3$  & $3.8\times10^{-5}$  \\ 
    $30$ &$10^{-3}$  & $2.1$  & $1.0\times10^{-6}$ \\ 
    $60$ &$10^{-3}$  & $3.5$  & $9.7\times10^{-7}$ \\ 
    $100$ &$10^{-3}$  & $5.3$  & $9.8\times10^{-7}$ \\ \hline
  \end{tabular}
\end{center}
\label{assumptions}
\end{table}

\textit{On the validity of the assumptions}. There are two assumptions that limit the cases which can be solved with the current model. The first one is the lubrication approximation, which restricts the usage of small corner radius and high superheat, since both $\delta/R_o$ ratio and Reynolds number increase with reduced corner radius and elevated superheat. As the superheat decreases, the film thickness and Reynolds number also decrease, but this time the second assumption, continuum, becomes a limiting parameter, since the liquid film at the corner may become extremely thin. Table~\ref{assumptions} summarizes the parameters related to these two assumptions for the selected cases. For $30\unit{nm}$ corner radius, the time scale ratio decreases from the order of $10^{-2}$ to $10^{-6}$ as the superheat decreases from $1.0\unit{K}$ to $10^{-3}\unit{K}$. Keeping the superheat at $10^{-3}\unit{K}$, the time scale ratio is of same order for 30, 60 and 100$\unit{nm}$ corner radius. However, the minimum film thickness is much smaller for the sharper corner. Therefore, the radius of curvature of the corner is limited by the lubrication assumption for high superheats and is limited by continuum assumption for low superheats. It is worth emphasizing that for the case with $10^{-3}\unit{K}$ superheat and $30\unit{nm}$ corner radius, the minimum local film thickness becomes $2.1\unit{nm}$, which may necessitate the inclusion of non-continuum effects, such as wall slip and molecular layering. However, the region of this extremely thin film occurs only around the local minima of the film and the extent of this region is highly restricted.

\section{Conclusion}
\label{sec:conclusion}

A solution methodology is developed to construct a comprehensive model of thin film condensation in a fin-groove system, which is the representative unit structure for grooved wicks. The proposed model enables the investigation of the effect of dispersion forces (without structural effects) on the film profile of a non-polar liquid by keeping the fin-groove corner, where the film becomes thinnest, inside the solution domain. The results demonstrate that reducing superheat and sharpening of the fin-groove corner promote the effect of the disjoining pressure on the thin film profile. When the dispersion forces are effective in the corner region---which corresponds to the case with $30\unit{nm}$ corner radius and $10^{-3}\unit{K}$ superheat in the present study---the liquid film conforms to the solid surface in the close proximity of the corner leading to an abrupt change, a \textit{slope break}, in the film profile. A similar finding was reported in a previous experimental study \cite{lips2010}, where the axial liquid flow and disjoining pressure were speculated to be the possible sources for the \textit{slope break}. Although the problem modeled in the current study is not identical to the problem experimented, a similar configuration (without axial flow) studied in the current work reveals that disjoining pressure is able to bend the liquid-vapor interface such that a thinner film forms on the fin top, which leads to a higher condensation rate due to the decreased thermal resistance.  Future research is intended to focus on the effects of polarity and structural forces such as molecular layering in a similar fin-groove system.

\addcontentsline{toc}{section}{References}
\bibliographystyle{unsrt}
\bibliography{references}

\section*{Declarations of interest}
\addcontentsline{toc}{section}{Competing_interests}
None.

\end{document}